\documentstyle[11pt,epsfig,amssymb,citesort]{article}
\def\parin{\hskip \parindent }

\newcommand{\beq}{\begin{eqnarray}}
\newcommand{\eeq}{\end{eqnarray}}
\newcommand{\eq}{\ref}

\newcommand{\Lqcd}{\Lambda_{{\rm QCD}}}

\newcommand{\be}{\begin{equation}}
\newcommand{\ee}{\end{equation}}
\newcommand{\lwrsim}{\raise0.3ex\hbox{$<$\kern-0.75em\raise-1.1ex\hbox{$\sim$}}}
\newcommand{\mucinq}{\gamma_\mu\gamma_5}
\newcommand{\pluscinq}{\gamma^+\gamma_5}

\def\C2#1#2{({\cal C}_2)_{#1}^{#2}}

\def\eqti{ {\,\,\lower.7ex\hbox{$=\atop t\rightarrow\infty$}\,\,}}
\def\propti{ {\,\,\lower.7ex\hbox{$\propto\atop t\rightarrow\infty$}\,\,}}
\def\sub#1{\underline{#1}}
\def\eq#1{eq. (\ref{#1})}

\def\ident{{ \mathchoice {1\mskip-4mu\mathrm{l} }
{1\mskip-4mu\mathrm{l} }
{1\mskip-4.5mu\mathrm{l} } {1\mskip-5mu\mathrm{l}} }}

\begin{document}
\setcounter{page}{1}
\begin{flushright}
LPT-ORSAY 00/111\\
UHU-FT/00-03\\
\end{flushright}
\begin{center}
\bf{\huge 
Preliminaries on a Lattice Analysis of The Pion Light-cone Wave function: 
a Partonic Signal?}
\end{center}  
\vskip 0.8cm
\begin{center}{\bf A. Abada$^a$, Ph. Boucaud$^a$, 
G. Herdoiza$^a$,
 J.P. Leroy$^a$,\\  J. Micheli$^a$,
O. P\`ene$^a$, J. Rodr\'\i guez--Quintero$^b$
}\\
\date={\today}
\vskip 0.5cm 
$^{a}$ {\sl Laboratoire de Physique Th\'eorique~\footnote{Unit\'e Mixte 
de Recherche du CNRS - UMR 8627}\\
Universit\'e de Paris XI, B\^atiment 211, 91405 Orsay Cedex,
France}\\
$^b${\sl Dpto. de F\'{\i}sica Aplicada e Ingenier\'{\i}a el\'ectrica \\
E.P.S. La R\'abida, Universidad de Huelva, 21819 Palos de la fra., Spain} \\

\end{center}
\begin{abstract}

We present the first attempt of a new method to compute the pion light-cone wave
function (LCWF) on the lattice.   We compute the matrix element between the pion and
the vacuum of a non-local operator: the propagator of a ``scalar quark'' 
(named for short ``squark"). 
A theoretical analysis shows that for some kinematical conditions  (energetic pion and
hard squark) this matrix element depends dominantly on the LCWF $\Phi_\pi(u), u\in
[0,1]$.  On the lattice, the discretization of the parton momenta 
imposes further constraints on the pion momentum.  The two-point Green 
functions made of
squark-quark and squark-squark fields  show hadron-like bound-state behaviour
 and verify the
standard energy spectrum. We show some indications that during a short time,
after being created,  the system
of  the spectator quark and the squark behave like partons, before they form a
hadron-like bound state. This short time is the place where the partonic wave function
has to be looked for.

\end{abstract}
\date={\today}\newpage
\section{Introduction}
\label{sec:intro}

\parin The light cone wave functions (LCWF)~\footnote{We use the expression  ``light
cone'' wave function according to a common habit although ``null plane'' wave
function is more appropriate since the quantification surface is indeed a null
plane.} \cite{brodsky}  enter the calculation of a large variety of processes
such as electroweak decays, diffractive processes, meson production in $e^+e^-$
and $\gamma \gamma$ annihilation, relativistic heavy ion collisions, heavy
flavors and many others \cite{stan1}.

The LCWF depends on a large momentum scale, $\mu^2$, which is typically the
momentum of the considered hadron $P_z^2$ in a physically well chosen reference frame
(e.g. equal velocity frame for form factors,  $B$ rest frame for $B$
decay, etc.). The pion wave function is expanded in terms of Fock states:
\begin{equation} |\pi \rangle= a_1 |q\bar {q}\rangle + 
a_2 |q\bar {q}g\rangle +
a_3 |q\bar {q}gg\rangle + \cdots\, .\label{Fock} \end{equation}
 where the lowest Fock state
$|q\bar {q}\rangle$ describes the valence configuration which is dominant at
large enough $P_z^2$ \cite{farrar}. 
Up to power corrections $O(\Lambda_{\rm QCD}^2/P_z^2)$, the valence  component 
$|q\bar {q}\rangle$ is fully described by its leading twist amplitude. 

The leading twist amplitude 
has been proven to be describable in a very compact and frame independent way:
 the wave function
$\Phi_\pi(u)$ is defined by the  following matrix element involving the
$\pi^-$-meson and a light cone Wilson string \beq \langle 0| \bar d(0){\cal
{P}}\left[\ \exp(i\int_x^0 \ d\tau_\mu A^\mu)  \right]\mucinq u(x)
|\pi^-(p)\rangle_{x^2=0} =  -i p_\mu f_\pi \int_0^1 du\, e^{-iup\cdot x} 
\Phi_\pi(u) \; \; . \label{fdonde} \eeq  
 The Wilson string
in the square bracket ensures  the gauge invariance of the l.h.s. of eq.
(\ref{fdonde}). The link between the first term in eq. (\ref{Fock}) and $\Phi_\pi(u)$
 will be dicussed in subsection \ref{derivation}.

Let us notice here that eq. (\ref{fdonde}) describes the
LCWF a la Bethe Salpeter (BS), but, although the Bethe Salpeter framework differs
significantly from the null-plane quantization approach, eq.
(\ref{fdonde}) exactly describes the dominant contribution to the pion wave 
function on a null plane. It is also useful to remind
that the null-plane quantized wave function on a plane $t+z=0$, is equal to 
the pion wave function quantized on $t=$ constant, for a pion with a
 momentum $P_z=\infty$. In eq. (\ref{fdonde}) $u$ denotes the longitudinal
momentum fraction of the  pion carried by the (valence) quark in the infinite
momentum frame.  The antiquark carries a  fraction ($1 ~- ~u$).

Let us insist, the pion wave function in QCD is an
extremely complicated object, which cannot be reduced to the 
BS wave function on the light cone~\footnote{The light cone is a surface of zero
measure in full space time.}. 
 However, in its {\it infinite momentum frame},  
  it simplifies dramatically in the following sense: the form factors 
  depend only on the longitudinal wave function defined in  eq. (\ref{fdonde})
 while the transverse motion of quarks  becomes irrelevant.    
For finite but large pion momenta the corrections
are $O(\Lambda_{\rm QCD}^2/P_z^2)$.  Equivalently, for a quark and an
 antiquark lying almost on the same light line a corrective term 
 $O(x^2\Lambda_{\rm QCD}^2)$ has to be added to the l.h.s. of eq. (\ref{fdonde}) if 
this is not to be  restricted to $x^2=0$.

Systematic expansions in inverse powers $\Lambda_{\rm QCD}^2/P_z^2$
may be performed. But, even better, for  each order in $\Lambda_{\rm QCD}^2/P_z^2$,
 perturbative QCD 
(pQCD) methods \cite{brodsky,efremov,bertsch} allow the coefficients to be 
systematically expanded  
in powers of $1/\log(P_z^2/\Lambda_{\rm QCD}^2)$. 
 
The dominant term in this perturbative expansion, i.e. the asymptotic form of the 
LCWF for very large $\mu^2\sim P_z^2$ reads:
\beq
\Phi^{\rm as}_\pi(u) = 6 u (1-u) \label{asympt}
\eeq
In this extreme limit, the shape of the wave function is totally given by 
pQCD, while the multiplicative constant $f_\pi$ in eq. (\ref{fdonde}) contains 
 all the relevant  non-perturbative knowledge.
The function (\ref{asympt}) is corrected by terms which decrease 
only logarithmically when $\mu^2\to \infty$. While the anomalous dimensions
of these terms  are computable from pQCD, their coefficients
are only computable by non perturbative methods or to be taken
from experiment.

At lower $\mu^2$, when the $O(\Lambda_{\rm QCD}^2/\mu^2)$ power corrections can
still be neglected but not the logarithmic  $O(1/\log(\mu^2/\Lambda_{\rm QCD}^2))$
ones,  the form of the wave function evolves away from eq (\ref{asympt}). 
The study  of the LCWF in this range needs the use of non-perturbative
methods. Most frequently one computes the LCWF via moments of the function
$\Phi_\pi(u)$ as will be shortly described in the next paragraph.
A well known example is the work by Chernyak and Zhitnitsky (CZ) 
\cite{CZ} who used the 
QCD sum rules~\footnote{These QCD sum rules for the first two moments of
the pion twist-two distribution amplitude were recalculated in ref.~\cite{BF}
 resulting in a shape between the two extreme cases $\Phi^{\rm as}$ and 
 $\Phi^{\rm CZ}$.} to determine the first two moments and obtained that
at $\mu= 1$~GeV the shape of the pion wave function is completely different from
its asymptotic form and it writes:
\beq
\Phi^{\rm CZ}_\pi(u) = 120 u (1-u)(u-0.5)^2 \label{eq:CZ}
\eeq
As can be seen from eqs. (\ref{asympt}) and (\ref{eq:CZ})
 there is a large difference between the two functions.

Experimental measurements of the electromagnetic form factors of the pion were
considered to be the best
way to study these wave functions \cite{stesto}.
 Recent model-dependent analyses of CLEO data on meson-photon
transition form factors \cite{cleo,rad} are consistent with the asymptotic
wave function. 
A direct measurement \cite{E791}  was carried out using data on
diffractive dissociation of 500 GeV/c $\pi^-$ into di-jets from a platinum
target at Fermilab experiment E791. The results show that the 
 asymptotic wave function (\ref{asympt}) describes the data well for $\mu^2
\sim 10 ~{\rm (GeV/c)^2}$ or more, although this interpretation
is subject to some controversy \cite{THE791}.

On the theoretical side, a direct  non perturbative measurement 
of the LCWF is badly wanted. There are only few  attempts in that direction.
The first method \cite{sac1} is a lattice computation of moments of 
the LCWF
\beq \label{eq:phidef} {\cal M}_n = \int_0^1 du \ u^n \Phi_\pi(u) \, . \eeq
 which can be done computing the pion to vacuum 
matrix elements of local operators such as  
\beq 
\langle  \pi^-(\vec p_\pi) \vert \bar d(0)\gamma^\mu \gamma_5
(i D^{\mu_1}) \dots
(i D^{\mu_n}) u(0)  \vert 0 \rangle &=& -i f_\pi {\cal M}_{n}
 p_\pi^\mu p_\pi^{\mu_1}
\dots p_\pi^{\mu_n} \nonumber + \dots  \label{operators}
\eeq
where the dots at the end correspond to terms suppressed by powers of 
$\Lqcd^2/P_z^2$ (the same terms have been eliminated in 
the l.h.s. of Eq. (\ref{fdonde}) by means of the restriction $x^2=0$). 
The lattice discretization of the derivative operators 
in (\ref{operators}) is more and more tricky with higher moments, and
their renormalization isn't easy either.

It was therefore proposed in \cite{aglieti} to attempt a direct 
calculation of the LCWF from lattice QCD. One tries to
``see'' on the lattice {\it the partonic constituents } of the hadrons
instead of the hadrons themselves. The idea is first to consider an 
energetic pion, which is supposed to have its partonic constituents
``frozen'' by Lorentz boost, and second to 
 hit one of its quarks by giving it a large momentum in order 
 to measure the perturbative  part
  (small distance between the constituents) of the
  wave function. 
 A scalar with the color content of quarks
  propagating from the hit quark to the spectator one
 insures gauge invariance.

In this paper we report the first and preliminary real attempt in that
direction. In  section \ref{principle} we explain the principle 
of the calculation and derive the basic formulae, with a particular care
at establishing for which parameters of the run we may expect 
the subdominant contributions to the pion wave function to be
under control. In section \ref{lattice} we describe the lattice
 set-up which was used. 
 In section \ref{twopoint} we present the results
 on the two-point Green functions. 
 In section \ref{sec:threepoint} we present the results on the three-point Green
 function and present the main analysis of our result. We believe that
 our results might provide some hint of a partonic behaviour. 
Finally, we discuss the relevance of our results in section \ref{conclusion}.

\section{Principle of the calculation}
\label{principle}

\parin In this section we want to elaborate some theoretical tools necessary
to prepare the direct lattice calculation of the LCWF. The issue is to
reach some understanding of what to run on a lattice 
to measure the pion LCWF and to estimate the expected uncertainties.
It is clearly impossible on a lattice to measure directly the matrix element
in eq. (\ref{fdonde}) since obviously Euclidean metric has no light cone.
The large momentum frame approach is more promising, with a standard
continuation to imaginary time. We will then need to take into consideration 
the full pion wave function, assuming from QCD some general knowledge
about it, and then consider under which conditions what is measured
in the lattice depends dominantly on the LCWF, and if so, to estimate the
subleading contributions. This will first be performed
in Lorentz metric in an infinite volume. Later on we will take into
account the Euclidean metric and the finite volume effects.

\subsection{Derivation of the basic formulae}
\label{derivation}

\parin From now on, we will  use the Light-cone gauge, where the path ordered
operator\\  ${\cal {P}}\ \exp(i\int_x^0 \ d\tau_\mu A^\mu)$ is equal to $1$.
Equation (\ref{fdonde}) defines the  pion Bethe \& Salpeter  wave function   on
the light cone, which has been extensively studied in literature since the
pioneering work of Brodski and Lepage \cite{brodsky}. It contains  the leading 
contribution to the pion wave  function, the subleading pieces having been
eliminated by the light cone condition $x^2=0$.  We are aiming at a lattice
investigation of this wave function.  This will lead us (as already done in ref.
\cite{aglieti}) to compute Fourier integrals of the wave function over  the
whole  space and not only on the light cone.  Therefore, the effect of
subdominant contributions should be considered.   Luckily, hadron properties, as
derived from QCD asymptotic freedom, allow  to control the approximation
introduced  when neglecting these subdominant contributions. 

 Let us follow the standard Light-cone perturbation theory (LCPth) techniques
\cite{brodsky}. We consider the first term in eq. (\ref{Fock}), i.e.  the valence $\bar{u}-d$ Fock
state~\footnote {Strictly speaking, we retain only the dominant part of the valence Fock state,
the one connected to vacuum via the axial current, the other  contributions being suppressed. 
This suppression 
can be understood simply from the fact that the quarks in an energetic pion have dominantly the same
helicity.} for the $\pi^-$-meson wave function resulting from the quantification on the null-plane
time {\it i.e.} $x^+=t+z=0$\\  ($V^{+(-)}=V_0+(-)V_z$):
\[ <0| \bar d(0) \pluscinq u( \ \sub{x} \ ) |\pi( \ \underline{p} \ )>_{x^+=0} =  -i p^+
f_\pi \int_0^1 du\,e^{-iu\frac{p^+x^-}{2}} \int \frac{d^2\vec{k}_\perp}{(2\pi)^2}
\,  e^{i\vec{k}_\perp \cdot\vec{x}_\perp}  \psi_{\bar u d/\pi}(u,\vec{k}_\perp) \]
\beq = -i p^+
f_\pi \int_0^1 du\,e^{-iu\frac{p^+x^-}{2}}\,\widetilde{\psi}_{\bar u d/\pi}
(u,\vec{x}_\perp) \; \; 
 \label{fdondecomp}\eeq 
where the change of variable $k^+=up^+$ has been performed, with $0\le u\le 1$
since both the ``+'' components of quark ($up^+$) and antiquark 
($(1-u)p^+$) have to be positive (remember that components ``+'' of momenta have 
to be positive by definition) 
and where $\widetilde{\psi}_{\bar u d/\pi}(u,\vec{x}_\perp)$ is the partial
Fourier transform (over $\vec{k}_\perp$) of
 $\psi_{\bar u d/\pi}(u,\vec{k}_\perp)$.  

The previous matrix  element depends on the light-cone
three-momentum $\sub{p}=(p^+,\vec{p}_\perp)$ and its conjugated three-vector in
configuration space, $\sub{x}= (x^-,\vec{x}_\perp)$.  For the sake of simplicity,
we chose  the frame where  $\vec{p}_\perp=0$, and hence $p \cdot x \equiv
p^+x^-/2$.  The wave function $\psi_{\bar u d/\pi}(u,\vec{k}_\perp)$ in Eq.
(\ref{fdondecomp}) represents the probability amplitude for finding two partons
with momenta $(up^+,\vec{k}_\perp)$ and $(p^+(1-u),-\vec{k}_\perp)$ 
respectively in the
valence Fock state of the pion.  This amplitude is normalized to $1$, 
\beq \int_0^1 du \ \int \frac{d^2\vec{k}_\perp}{(2\pi)^2} \psi_{\bar u
d/\pi}(u,\vec{k}_\perp) = \int du \widetilde{\psi}_{\bar u d/\pi}(u,0)= 1 \ , \eeq
\noindent as it immediately comes from requiring that $<0| \bar d \pluscinq u
|\pi>= -ip^+ f_\pi$ when the operator becomes local, i.e. when $\sub{x}=0$ in Eq.
(\ref{fdondecomp}).

In eq. (\ref{fdondecomp}) we have only considered the $\gamma_\mu$ component in
the direction  ``+'' of the pion momentum. The other directions $\gamma^\perp$
and $\gamma^-$ can lead to matrix elements proportional respectively to $p^\perp$
and $p^-$. In the pion rest frame all these components of the matrix elements
should be of order $\Lambda_{\rm QCD}^2$ if we do not assume any
restriction~\footnote{Let us repeat that we are not allowed to restrict ourselves
to small $x_\mu$ since we will perform Fourier transforms.} on $x$. This  simply
expresses that the size of the pion in its rest frame is $O(\Lambda_{\rm QCD})$
in momentum space and $O(1/\Lambda_{\rm QCD})$ in configuration space. Let us now
consider a frame in which the pion has a very large $p^+$. Then the matrix
element  considered in eq. (\ref{fdondecomp}) is increased proportionally to the
increase of $p^+$, on the contrary $x^-$ is decreased by the same ratio and the
transverse components stay constant. For an ``infinite momentum'' pion we are
left only with the contribution proportional to the pion momentum. This is a
first indication that in our analysis we will  have to concentrate on energetic
pions.

Equation. (\ref{fdondecomp}) is a definition of 
the wave function $\psi_{\bar u d/\pi}(u,\vec{k}_\perp)$.
It only depends on the
quantities $u$ (the fraction of pion's momentum carried longitudinally by one
parton) and $\vec{k}_\perp$; it is frame-independent for longitudinal boosts. 
 In order to establish the connection with eq. (\ref{fdonde}), we now put $x^2=0$
  ({\it i.e.}
$\vec{x}_\perp=0$ provided that we quantized on the light-cone time $x^+=0$) in
eq. (\ref{fdondecomp}). If we take $\vec{x}_\perp=0$ in eq. (\ref{fdondecomp})
and compare the result with eq. (\ref{fdonde}) we see that~\footnote{
Remember that the exponential in brackets in eq. (\ref{fdonde}) is equal to 1 in
our gauge.} 

\beq \Phi_\pi(u) \equiv \int \frac{d^2\vec{k}_\perp}{(2\pi)^2} \psi_{\bar u
d/\pi}(u,\vec{k}_\perp) \, = \widetilde{\psi}_{\bar u d/\pi}
(u,0) \;\; . \label{phipi} \eeq

To clarify the physical picture let us now compare in the Light-cone gauge
the l.h.s of eq. (\ref{fdonde}) unrestricted to $x^2=0$ (the full BS equation)
 and the l.h.s of eq. (\ref{fdondecomp}).
 They only differ by the  null plan constraint $x^+=0$. 
This constraint is generated by requiring that  
the pion carries a large momentum.  
Indeed $p^- = m_\pi^2/(p_z+E_\pi)$ appears to be
powerly suppressed. 
\noindent This suppression of $p^-$ implies that 
  $p^+x_- + p^- x_+ \simeq p^+ x_-$ (unless $x^+$ is unnaturally large).
If one assumes the absence of sudden changes when $x^+$ moves away from 0, one may replace 
$p^+ x_-$ by $p \ x$ in eq. (\ref{fdondecomp}) which now reads:
\beq <0| \bar d(0) \mucinq u(x) |\pi^-(p)> =  -i p_\mu f_\pi \int_0^1 du\,
e^{-iup\cdot x}  \widetilde{\psi}_{\bar u d/\pi}(u,\vec{x}_\perp) \; \; ,
\label{fdondef} \eeq
 If we add the physical input that the  
wave function extends typically to transverse momenta of the order of $\Lqcd$, 
we get from $0\le u\le 1$ the picture that the valence
constituents of the pion move essentially in the same direction than the pion
itself at a velocity close to 1. In other words, due to asymptotic freedom,
the constituents do not like to have a very large virtuality and the only way
for almost massless quarks to build up the energy and momentum of the 
almost massless pion is to move in the same direction, i.e. to have 
$E_q + E_{\bar q}\simeq |k_q| +
 |k_{\bar q}| \simeq E_\pi \simeq |k_q + k_{\bar q}|$

 From now on we shall follow the method in \cite{aglieti} and we will 
 replace the gauge  invariance restoring  operator 
${\cal {P}}\ \exp(i\int_x^0 \ d\tau_\mu A^\mu)$, by another one, which is
 easier to continue analytically to euclidean time:  
the scalar coloured propagator,
\beq S(0;x)&=&\frac 1{-{\cal D}^2 - m_S^2 + i\epsilon}\simeq 
\frac {1}{-{\cal \partial}^2 
- m_S^2
+ i\epsilon} \nonumber \\ &=&\int \frac{d^4k}{(2 \pi)^4} e^{- ik\cdot (0-x)} 
 \frac i{k^2- m_S^2 + i\epsilon} \; \; . \label{prop} \eeq
 where $m_S$ is a mass parameter, assumed to be small or zero to
mimic a  massless parton. In eq. (\ref{prop}), when replacing ${\cal D}^2$  by
${\cal \partial}^2$ we have bluntly neglected the coupling to gluons. This has
been done in order to simplify the argument which will follow and is justified if
we assume the scalar object to be ``hard'' and hence to behave mainly as a
parton.  Still,  a careful study of the effect of radiative corrections is
strongly needed.
This replacement looses the gauge invariance of the $1/{\cal D}^2$ operator. 
This is difficult to avoid: if the light cone wave function (\ref{fdonde}) is
gauge invariant,  the more general ones, eq. (\ref{fdondef}) are not. The loss of
gauge invariance is here the price we pay to present the argument which will
follow. Needless to say,  the real lattice calculations have been performed in a
gauge invariant way.

\noindent We can thus write

\beq
 & & e^{-iq\cdot x}\,<0| \bar d(0) \mucinq S(0;x) u(x) |\pi(p)> \nonumber \\
 &=&
 -i p_\mu f_\pi \int_0^1 du 
\int \frac{d^4k}{(2\pi)^4} e^{-i(up + q-k )\cdot x}
\frac i{k^2- m_S^2 + i\epsilon} \widetilde\psi_{\bar u d/\pi}(u,\vec{x}_\perp) 
 \eeq

\noindent This is supposed to be valid for all $x$
 so that we can integrate over $\vec{x}$ and obtain:

\beq & & i\int d^3x \  e^{-iq\cdot x}\,<0| \bar d(0) \mucinq S(0;x) u(x) |\pi(p)>
\nonumber \\ 
&=& p_\mu f_\pi \int_0^1 du \int \frac{dk_0}{2\pi}\, dk_z \,\frac{d^2
\vec{k}_\perp}{(2\pi)^2} \, e^{i(u E_\pi -k_0 )\, t}\, \frac {i \delta(up_z+q_z-k_z)
}{\vec{k}_{//}^2 - \vec{k}^2_\perp - m_S^2 + i\epsilon} \psi_{\bar u
d/\pi}(u,\vec{q}_\perp-\vec{k}_\perp) \nonumber \\ 
&=&-\, p_\mu f_\pi \int_0^1 du \int
\,\frac{d^2\vec{k}_\perp}{(2\pi)^2} \, \frac {e^{i\left(u E_\pi -
\sqrt{(up_z+q_z)^2+(\vec{q}_\perp+\vec{k}_\perp)^2  + m_S^2}\,\right)\,t} } {2
\sqrt{(up_z+q_z)^2+(\vec{q}_\perp+\vec{k}_\perp)^2 + m_S^2}}\, \,
\psi_{\bar u d/\pi}(u, -\vec{k}_\perp) 
\eeq

\noindent where $q_0=0$, $x_0=-t$ ($t<0$), $\vec{k}_{//}=(k_0,k_z)$ and again
$\vec{p}_\perp=0$. The r.h.s. of the latter line derives from integrating the former's
over $\vec{k}_{//}$ and changing variables  $(q-k)_\perp \to -k_\perp$. 

At this stage let us return to the physical understanding of the
wave function $\psi_{\bar u d/\pi}(u, -\vec{k}_\perp)$
already briefly considered above. The quarks  have a small probability
  of being far off shell  and $\psi_{\bar u d/\pi}(u, \vec{k}_\perp)$
 vanishes when   $\vec{k}_\perp^2$ becomes large~\footnote{Perturbative
analysis indicates that hadronic wave functions do not decrease quickly enough as
$\vec{k}_\perp^2 \to \infty$ to avoid the appearance of infinities. The Pion
 $\bar q q$-wave function falls off roughly as 
$1/\vec{k}_\perp^2$~\cite{brodsky},
and the resulting UV logarithmic divergence is the origin of the scale dependence
of the wave function.
For the sake of simplicity this point shall be deliberately overlooked in our 
formal derivation.}. In practice,  $\vec{k}_\perp^2 \psi_{\bar u d/\pi}(u,
\vec{k}_\perp) \to 0$ as  $\vec{k}_\perp^2 \to \infty$~\cite{brodsky}. Therefore, this
suppression for large $\vec{k}_\perp^2$ allows one to expand in powers of the
transverse components, provided that $E_S \gg \Lambda_{\rm{QCD}}$;
$\Lambda_{\rm{QCD}}$ being a natural hadronic energy scale {\it bounding} the
 transverse momentum carried by the partons and 
\beq
E_S\equiv\sqrt{(up_z+q_z)^2+ q^2_\perp + m_S^2}.\label{ES}
\eeq 
We then get:
\beq 
& & i\int d^3x \  e^{-iq\cdot x}\,<0| \bar d(0) \mucinq S(0;x) u(x) |\pi(p)> \nonumber \\
&=& -\,p_\mu f_\pi \int_0^1 du \,\frac{ e^{i (u E_\pi - E_S)\,t}}{2 E_S} \,  
\times  \Bigg[ \, \, \Phi_\pi(u) \nonumber \\
&+& \int \, \frac{d^2\vec{k}_\perp}{(2\pi)^2} \left\{ 
e^{-i \frac{2 \vec{k}_\perp \cdot \vec{q}_\perp 
+ \vec{k}_\perp^2}{2 E_S}\,t} \,\left( 1 \, - \,
\frac{2 \vec{k}_\perp \cdot \vec{q}_\perp + \vec{k}_\perp^2 }{2 E_S^2}
\right)\, - \, 1 \right\} \,\psi_{\bar u d/\pi}(u, -\vec{k}_\perp) \Bigg]
+ ... \; , \nonumber \\ \label{LightConeMin}
\eeq
It is easy to see that the second term inside the bracket (last line) is
formally  $O(\Lambda_{\rm QCD}/E_S)$, provided that $t \ll \,
E_S/\Lambda_{\rm{QCD}}^2$ and $t \ll \,E_S/(\Lqcd |\vec{q}_\perp|)$. This second
term is negligible as long as, and this is the general situation, $E_S\sim
[|up_z+q_z|^2+q_\perp^2]^{1/2}\gg \Lqcd~$\footnote{remember that $m_S$ is
small.}. However, when $|up_z+q_z|\sim \Lqcd $ for some values of $u$ and when 
$q_\perp \lwrsim \Lqcd$, i.e. when  $\vec p$ and $\vec q$ are back to 
back~\footnote{Strictly speaking $\vec p$ and $\vec q$ could be back to back
as long as $|q_z| - |p_z| \gg \Lqcd$.}, the
expansion  in eq. (\ref{LightConeMin}) breaks down as $E_S$ is not 
larger than $\Lambda_{\rm QCD}$ any longer. 
In another language, giving a large transverse kick to the pion generates a
hard gluon exchange between quarks which selects the perturbative component
 of the pion wave function, the so-called ``small pion'', which is what we 
 want to measure. Indeed in the FNAL experiment E791 \cite{E791}, the LCWF
 is observed via jets which have rather large transverse momenta. Let us now summarize. 

\paragraph{Conditions for a partonic signal:}[C1] In order to reach some knowledge  on the light-cone wave
function $\Phi_\pi(u)$ from the lattice calculation of the l.h.s of eq.
(\ref{LightConeMin}), the following conditions are required beyond the general 
large pion momentum constraint i.e. $p_z \gg \Lqcd$ : 
$t\ll \, E_S/\Lambda_{\rm{QCD}}^2$,  $t \ll  \, E_S/(\Lqcd |\vec{q}_\perp|)$ and
$E_S \gg \Lqcd$ for all $u$. This generally implies  $\cos_{\rm min}
\lwrsim \cos \theta_{pq}$ for some $\cos_{\rm min}$ significantly greater
than -1.

\subsection{Consequences of discrete partonic momenta.}
\label{discret}

\parin Let us now consider a finite  parallelepipedic volume with 
periodic boundary  conditions (torus).
As is well known, the momenta components can only take the form
\beq
p_\mu =\frac {2\pi}{L_\mu} n_\mu\label{pmu}
\eeq
where $n_\mu$ are integers and $L_\mu$ is the length in
the direction $\mu$.  This is obviously valid also for partonic
momenta~\footnote{For other values the amplitudes are canceled by 
destructive inteferences.}. Thus in the formulae of subsection \ref {derivation}
 all integrals over $\int_0^1 du$ have to be replaced by discrete sums
 over the values of $u$ such that $up_\mu$ verifies eq. (\ref{pmu}). 
 
 Here comes immediately a problem. Let us assume for one moment 
 that the components of $p_\mu$ are all $0$ or $2\pi/L_\mu$. 
 Then only the values 
 $u=0,1$ are allowed. In any model the  LCWF
which is proportional to $u(1-u)$, (\ref{asympt}) and (\ref{eq:CZ}), 
  vanish for these values. The expected dominant behaviour at large momentum
  is vanishing in this case, and only subdominant effects can be observed.

The simplest situation,  the only one considered from now on, is
      when the pion momenta are aligned along one of the lattice
      spatial directions $\mu$. To allow values of $u$ that scan the 
      domain of variation $[0,1]$ densely~\footnote{The dominant contribution 
      to the  LCWF is only possible when all the $n_\mu, \mu=1,3$ are 0 or have a common 
 divisor, and at least one $n_\mu$ is larger than 1.} enough to provide a fair
      description  of the LCWF we should have :
  \beq
 p_\mu \gg \frac {2\pi}{L_\mu}.\label{condition}
 \eeq


  This condition [C2] has to be added to the set of conditions [C1] summarized 
 at the end of  subsection \ref {derivation} in the case of infinite volume. 
 Clearly this new one {\it is not} equivalent to the former ones since this
 one does depend on $L_\mu$ and disappears smoothly in the large volume limit.

\subsection{Strategy for lattice calculations}
\label{strategy}

\parin Following the method of \cite{aglieti} we compute on the lattice the
three point Green function
\beq
F^\mu(\vec p,\vec q; t)\equiv
\int d^3y\, d^3x\, e^{-i\vec q\cdot \vec x}\,e^{-i\vec p\cdot \vec y}\,
< 0| P_5(\vec y, t_\pi)u(\vec x,t)S(\vec x,t;0)
 \mucinq \bar d(0)|0>\,e^{E_\pi (t_\pi - t)}.
\label{threepoint}
\eeq

\noindent When all the conditions [C1] and [C2] are satisfied, and after performing 
 a Wick rotation to
euclidean metric, the l.h.s. of equation (\ref{threepoint}) verifies approximately 
the following proportionality
  in terms of the LCWF:
\beq
 F^\mu(\vec p,\vec q; t)\propto\,p_\mu f_\pi \sum_{u_i} \,\frac{ 
 e^{- ((1-u_i) E_\pi + E_S)\,t}}{2 E_S(u_i)}\,\Phi_\pi(u_i)\label{tpphi}
\eeq
where the $\sum_i$
extends over all values $0\le u_i \le 1$ such that $u_i\, p_\mu * L/(2\pi)$
are integers.
$\vec p$ is the momentum of the pion generated by the interpolating field 
$P_5(y) \equiv \bar d(y) \gamma_5 u(y)$, $\vec q$ is a momentum given  to one
valence quark of the pion. $E_S(u)$ is defined in (\ref{ES}).
  We have assumed $0 < t < t_\pi$.  The $e^{E_\pi (t_\pi - t)}$ takes into account 
  the propagation
of the pion between $t$ and $t_\pi$.   Of course  $t_\pi - t$ has been assumed to be large
enough to eliminate  the excited pseudoscalar states.    

Eq. (\ref{tpphi}) may be understood in a simple way : the time evolution
between $0$ and $t$ is the product of the propagators of two ``partons'', one
scalar parton of energy $E_S$ with a propagator proportional to $e^{-E_S
t}/(2E_S)$ and the spectator quark of energy $(1-u) E_\pi$. The scalar parton has
the color quantum numbers of a  quark. For convenience let us  call it a squark
although it has obviously nothing  to do with supersymmetry. The three-point
Green function in eq. (\ref{threepoint}) could also be used to estimate the form
factor for the transition between a pion and squark-quark bound state (which we call a
pionino, $\widetilde \pi$, to follow on the same metaphoric nomenclature). 
In such a case we would
take $t$ large enough for the ground state pionino to dominate:

\beq
F^\mu(\vec p,\vec q; t) \propti e^{- E_{\widetilde\pi} \,t}
 \label{threepoint2}
\eeq
where $E_{\widetilde\pi}$ is the pionino energy.
For small $t$, on the contrary, the excited states should add up
coherently in a complicated manner. The analysis presented in the preceding
subsection seems to indicate that this should boil down to a rather
simple partonic-like picture. In other words we expect a kind of 
hadron-parton duality to be at work for small $t$ which should allow
a partonic reading of our data. 
At this stage it is clear that we need to study, beyond the three-point
function in the l.h.s of eq. (\ref{threepoint}), the two point function
corresponding to the pionino interpolating field. 
 
An additional comment concerns the squark mass. In the preceding formulae we have
written a squark mass $m_S$ as a free parameter.  In order to gain the richest
possible information on the pion wave function, the 
renormalized squark mass has to be as light as possible. How to perform this ?  We
have chosen an approach based on an analogy with QCD hadrons. We  will vary the
bare squark mass down to when the algorithm to compute the squark propagator
stops converging,  which we take as an indication of possible zero modes. 

Finally, all things considered, we will have to make a systematic study of the
spectrum of all the colorless bound states constituted by quarks and squarks. 
It will turn out that in the quenched approximation nice exponential behaviours
 do indeed 
appear, signaling the existence of pioninos and squark-squark bound states 
(see Fig.  
\ref{fig:twopoint}),
and furthermore,  for non vanishing momenta, they follow the relativistic 
spectral law $E=\sqrt{m^2+p^2}$ (see Fig. \ref{fig:einstein}), or if one prefers the lattice one
(see eq. \ref{spectral}) below),  
which is not distinguishable from the former within our statistical errors.

\subsection{Symmetries}
\label{symmetries}

\parin Before turning to the actual calculation, it is useful to summarize which among the
two- and three-point Green functions we intend to compute should vanish
because of  QCD's discrete symmetries.  

In general all Green functions we consider are real in configuration space. 
Therefore they are real in momentum space 
if parity-even and imaginary if parity-odd. See table \ref{tab_sym}.

\begin{table}[htb]
\begin{center}
\begin{tabular}{|c|c|c|c|c|c|}
\hline
 operator  & $\gamma$ matrices & parity & real/im &time reversal & vanishes at 
 $\vec p=0$\\ \hline
squark-squark    & $\ident$ &   +   &  real& + & no \\
squark-quark  & $\ident$&   +   &  real& + & no\\
squark-quark  & $\gamma_0$ & +  & real  &   - & no\\
squark-quark  & $\gamma_i$ & -  & imag   &   + & yes\\
quark-quark  & $\gamma_5-\gamma_5$ &   + &real  &   + & no\\
quark-quark  & $\gamma_0\gamma_5-\gamma_5$ &   + &real  &   - & no\\
quark-quark  & $\gamma_i\gamma_5-\gamma_5$ &   - & imag  &   + & yes\\
three-point & $\gamma_0\gamma_5- \gamma_5$ &  + &real &  -  & no \\
three-point & $\gamma_i\gamma_5- \gamma_5$ &  - & imag &  +  & yes \\
\hline
\end{tabular} 
\caption{\small This table shows the symmetry properties of the Green functions.
By three-point we mean the Green function $F^\mu(\vec p,\vec q)$ defined in eq.
(\ref{threepoint}). The second column refers to the $\gamma$-matrices in the Green
function. For squark-quark, only one $\gamma$-matrix is traced with the quark
propagator. In the other cases we indicate the matrices on both ends of the
quark propagators. The third column refers to the spatial parity of the Green
function. The time reversal refers to the symmetry when $t\to -t$ (and
$t_\pi \to -t_\pi$ in the the three point case). We thus learn, for example, that the three
point with $\gamma_0\gamma_5 - \gamma_5$ vanishes at $t=0$ if $t_\pi=t_{\rm max}/2$.}
\label{tab_sym}
\end{center}
\end{table}
\vskip 5 truemm

\section{Lattice set-up}
\label{lattice}

\parin We consider a $16^3\times 40$ lattice at $\beta=6.0$ in the 
quenched approximation. 
The quarks are computed with the clover action with the coefficient $c_{sw}=1.769$. 
We have used 
for the spectator quark two values of the bare mass parameters:
$\kappa=0.1333$ and 0.1339, and  for the active one  
$\kappa=0.1339$. 

The squark propagator $D(x,0)$ verifies the equation
\beq
\left[\delta_{x,y} - \kappa_S \sum_\mu \left(
U_\mu(x)\delta_{x,y-\hat \mu} + U^\dagger_\mu(x-\hat\mu)
\delta_{x,y+\hat \mu}\right)\right]D(y,0)
= \delta_{x,0}  \label{propagateur}
\eeq

We compute the squark propagator with the bare mass 
parameter $\kappa_S=$ 0.1428, 0.1430, 0.1431.
Above $\kappa_S = 0.1431$ the convergence of the inverter becomes 
very long, which we take as a sign that we are close to the 
massless squark.

\begin{figure}[hbt]
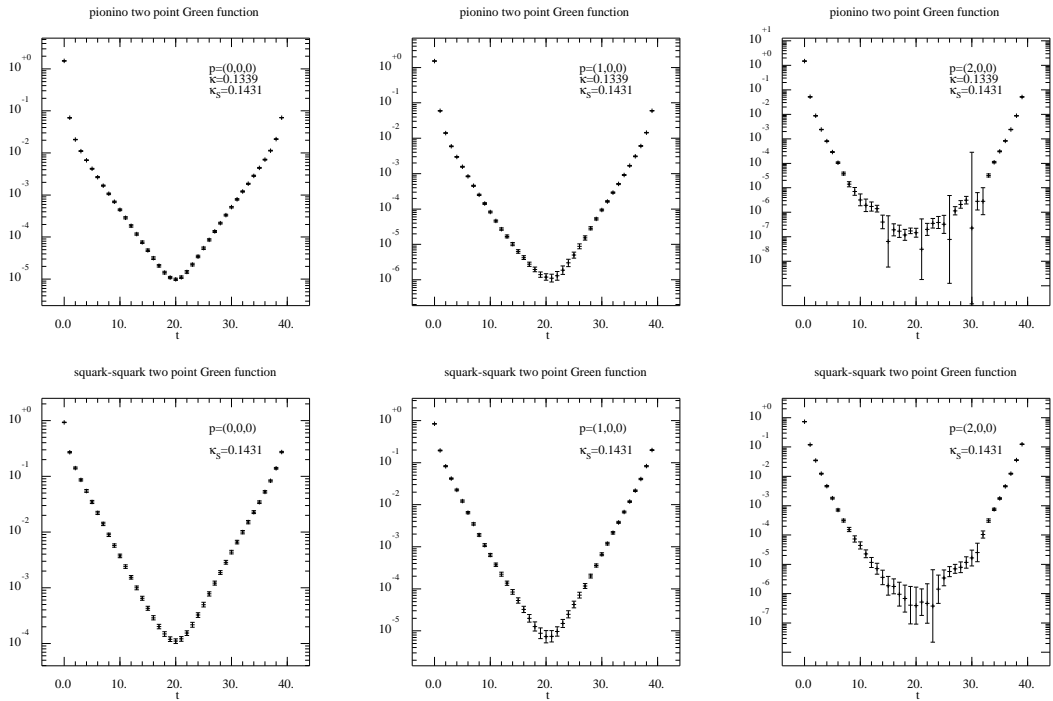

\hspace*{-1.8cm}
\begin{center}
\begin{tabular}{lll} 
\epsfxsize4.5cm\epsffile{pionino_0.eps} &
\epsfxsize4.5cm\epsffile{pionino_1.eps} & \epsfxsize4.5cm\epsffile{pionino_2.eps}\\
\epsfxsize4.5cm\epsffile{scasca_0.eps} & 
\epsfxsize4.5cm\epsffile{scasca_1.eps} &\epsfxsize4.5cm\epsffile{scasca_2.eps}\\
\end{tabular}
\end{center}
\caption{\small Two point Green functions in logarithmic plots
for pionino and squark-squark
states. As an example we present the lightest
states, i.e.  $\kappa=0.1339$ and $\kappa_S=0.1431$.}  
  \label{fig:twopoint}
  \end{figure}
  
In each case we have run 100 configurations. The errors are computed 
according to the jackknife method. The pion interpolating field
$P_5$ is inserted at $t_\pi = 16$. This has been chosen so that the 
direct signal at small $t$ is not significantly perturbed by the signal 
 which has looped around via the end of the lattice : 40-16 has eight 
 time-intervals more than 16. This is an important precaution. 
 Indeed from table \ref{tab_sym} we learn that the three-point Green
 function with $\gamma_0\gamma_5$ inserted at $t=0$ is odd for time reversal. 
 If $t_\pi$ was taken in the middle of the lattice, $t=20$, it 
 would have resulted a vanishing of this three-point Green
 function for $t=0$. Since we are interested in small values
 of $t$ such a vanishing of the signal would have made the analysis 
 impossible. 
  
  For the study of the two and three point Green functions 
we  have run with the following values for the pion three momentum:
\beq
\frac{L\vec p}{2\pi} = (0,0,0); (1,0,0);(1,1,0); (2,0,0)
\eeq
In practice, however, the vanishing momentum does not produce a pion describable
by a light-cone wave function.
 The momentum (1,0,0) (1,1,0)  will not be useful 
since in these cases only the values $u=0,1$ are allowed by the discretization 
 and  
the LCWF
 vanishes for these values. However, they are kept in the analysis for a comparison of the results 
 obtained from (1,0,0) (1,1,0)  with 
 the ones from (2,0,0), which  might be interpreted as partonic signal.

Concerning $q_\mu$ we have run a large number of momenta, with components
ranging from $-\frac {4 \pi}L$ to $\frac {4 \pi}L$ but again too large momenta
are too noisy. We will detail later the momentum configurations which
 are considered in the analysis.

 As already explained, we hope to catch the partonic signal 
 at small $t$. In practice we have concentrated on the region $t=0,4$ 
 as we will see later.
 It leaves $t_\pi-t \ge 12$ which should be enough to isolate
 the pion and it leaves some space to look for plateaus. 

\section{Two point Green functions}
 \label{twopoint}
 
\parin  We have shown in fig. \ref{fig:twopoint} six examples
 of new two point Green functions for momenta $\vec p = (0,0,0)$, 
 $\vec p = {2\pi/ L}(1,0,0)$  and 
 $\vec p = {2\pi/ L}(2,0,0)$ respectively. It is seen that these
 two point Green functions do indeed behave as if the quark-squark and
 squark-squark states were hadron-like bound state.

\begin{table} [hbt]
\begin{center}
\begin{tabular}{|c|c|c|c|c|}
\hline
 momentum & 0 & 1 & 1.4 & 2\\ \hline
pion $q_1q_1 - \gamma_5,\gamma_5$ &  0.42(2) & 0.62(3) & 0.70(4) &
0.61(13) \\ 
pion $q_1q_1 - \gamma_0\gamma_5,\gamma_0\gamma_5$ &  0.41(2) & 0.60(2) & 0.69(3) &
0.89(7) \\ 
pion $q_2q_1 - \gamma_5,\gamma_5$ &  0.38(2) & 0.60(3) & 0.66(4)  &
0.35(16) \\
pion $q_2q_2 - \gamma_5,\gamma_5$ &  0.34(2) & 0.58(4) & 0.61(5) &
0.09(22) \\
pion $q_2q_2 - \gamma_0\gamma_5,\gamma_0\gamma_5$ &  0.34(2) & 0.56(3) & 0.62(4) &
0.84(10) \\
rho $q_1q_1 - \gamma_i,\gamma_i$ &  0.62(1) & 0.76(2) & 0.96(3) &
1.02(6) \\
rho $q_2q_2 - \gamma_i,\gamma_i$ &  0.60(2) & 0.71(3) & 0.98(5) &
0.95(9) \\
pionino $q_1S1 - \gamma_0$ &  0.59(1) & 0.71(1) & 0.81(1) &
0.98(3) \\
pionino $q_1S1 - \ident$ &  0.55(1) & 0.67(1) & 0.77(2) &
0.91(5) \\
pionino $q_1S2 - \gamma_0$ &  0.54(1) & 0.67(1) & 0.77(2) &
0.95(3) \\
pionino $q_1S2 - \ident$ &  0.51(1) & 0.63(2) & 0.74(2) &
0.88(6) \\
pionino $q_1S3 - \gamma_0$ &  0.51(1) & 0.65(2) & 0.76(2) &
0.93(3) \\
pionino $q_1S3 - \ident$ &  0.48(2) & 0.60(2) & 0.72(2) &
0.86(7) \\
pionino $q_2S1 - \gamma_0$ &  0.57(1) & 0.70(1) & 0.79(1) &
0.98(3) \\
pionino $q_2S1 - \ident$ &  0.53(1) & 0.64(2) & 0.74(2) &
0.91(7) \\
pionino $q_2S2 - \gamma_0$ &  0.52(1) & 0.66(2) & 0.76(2) &
0.94(3) \\
pionino $q_2S2 - \ident$ &  0.48(1) & 0.60(2) & 0.71(3) &
0.87(9) \\
pionino $q_2S3 - \gamma_0$ &  0.49(2) & 0.63(2) & 0.74(2) &
0.92(4) \\
pionino $q_2S3 - \ident$ &  0.45(2) & 0.57(2) & 0.69(3) &
0.85(10) \\
squark-squark $S1S1$ &  0.59(2) & 0.70(2) & 0.80(2) &
0.93(5) \\
squark-squark $S2S2$ &  0.50(2) & 0.61(2) & 0.74(3) &
0.83(7) \\
squark-squark $S3S3$ &  0.44(2) & 0.56(3) & 0.72(4) &
0.74(8) \\

\hline
\end{tabular} \par
\vskip 5 truemm
\end{center}
\caption{\small Energies of the various bound states in units of $a^{-1}$
(for $\beta=6.0,\, a^{-1}\simeq 2.0$ GeV).
The symbols $q_1,q_2$ represent respectively $\kappa=0.1333,0.1339$ for quarks; 
$S1, S2, S3$ respectively $\kappa_S=0.1428, 0.1430, 0.1431$ for scalars.
The momentum norms are given in units of $2\pi/L$. 
 We indicate the 
$\gamma$ matrices used in the meson interpolating fields.}
\label{tab:deux_points}
\end{table}
\vskip 5 truemm

\begin{figure}[hbt]
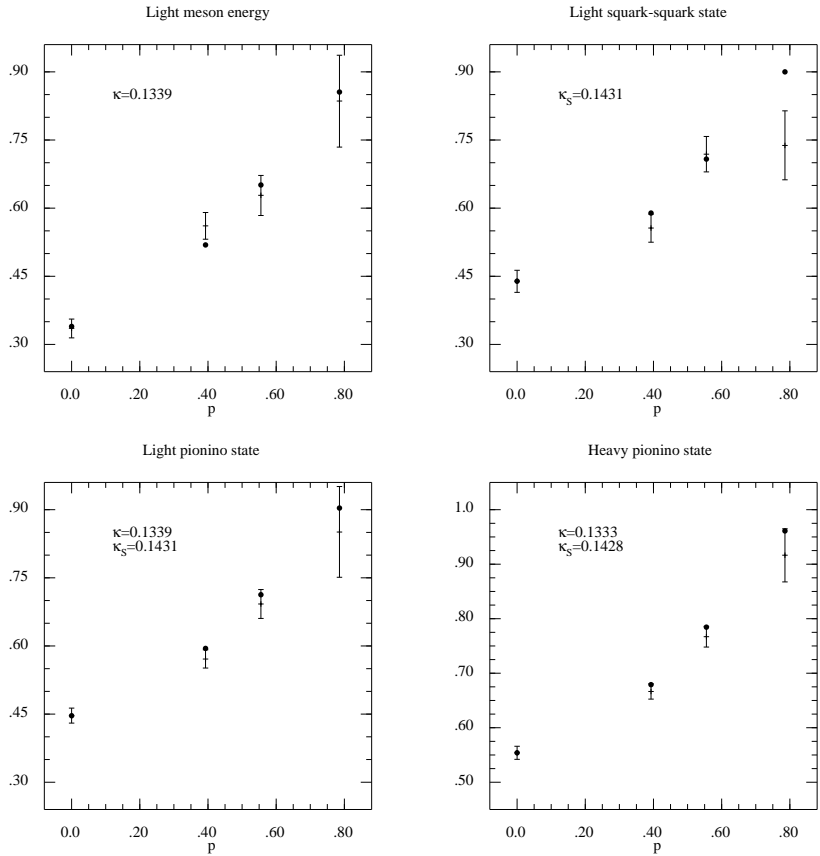

\hspace*{-1.3cm}
\begin{center}
\begin{tabular}{ll}
\epsfxsize5.5cm\epsffile{meson.eps} & 
\epsfxsize5.5cm\epsffile{sca_sca.eps} \\
\epsfxsize5.5cm\epsffile{en_pionino.eps} & 
\epsfxsize5.5cm\epsffile{heavy_pionino.eps} \\
\end{tabular}
\end{center}
\caption{\small Energy of the bound states as a function of the momentum
in lattice units. The dots correspond to the continuum formula 
$E=\sqrt{m^2+p^2}$, the mass being taken as the 
central value of the zero momentum energy. Three first
plots are with $\kappa=0.1339, \kappa_S=0.1431$, the last one
with $\kappa=0.1333, \kappa_S=0.1428$.}  
  \label{fig:einstein}
  \end{figure}

We present the results for the energies of the 
bound states in table \ref{tab:deux_points}. In fig \ref{fig:einstein} 
we present some checks of the spectral law $E=\sqrt{m^2+p^2}$. 
The latticized free boson dispersion relation 
\beq
\sinh^2(E/2)= 
 \sinh^2(m/2) + \sum \sin^2(p_\mu/2))\label{spectral}\eeq
  does not significantly
 differ from the continuum one within our errors.
 For momentum $4\pi/L$ the quark-quark states are 
 in some cases meaningless due to the noise. 
 It is surprising that the non conventional states present a better signal
 for this large momentum. 
 
 Of course the main lesson of this analysis is that the non-conventional
 bound states, pioninos and squark-squark do behave exactly as real hadrons.
 We are not in a position to discuss the theoretical implications of this 
 fact neither make any statement about the existence of such bound states
 in a non-supersymmetric extension of QCD. 
 
 The lowest bare squark mass considered is $\kappa_S=0.1431$.  When $\kappa_S$ is
varied slightly above 0.1431, the scalar inverter does not  converge anymore.
This squark is coded $S3$ in table \ref{tab:deux_points} and we see that the
corresponding squark-squark bound state rest mass is about 0.44 in lattice unit,
i.e. about 900 MeV ($a^{-1}\simeq 2$ GeV for $\beta=6.0$),  not far from
the rho meson mass. It is rewarding that the mass of this squark-squark bound
state is rather light, as if the squark with an approximately vanishing
renormalized mass did indeed produce rather light bound states~\footnote{We do not
know of any symmetry which would impose a pion-like massless state for massless
squarks.}. We feel  encouraged  to treat indeed this squark as a light parton as
will be done soon.    

 \section{Three-point functions}
\label{sec:threepoint}

\parin With our set of momenta, only the momentum $\vec p_\pi= {2\pi /L}\ (2,0,0)$ gives a
non-vanishing~\footnote{Notice that the CZ wave function (\ref{eq:CZ}) vanishes for
$u=1/2$, its study 
 needs even larger momenta and will not be discussed in our analysis.} 
 $\Phi_\pi(u)$ for  discrete $u=1/2$. 
 Thus we will focus our analysis on 
the latter momentum although  we have studied the full set of momenta $\vec p_\pi$,
 with a set of momenta $\vec q$ to be discussed later. We have only
 considered  the time  component ${F^0(\vec p,\vec q; t)}$. 
 
   Our analysis
of the data follows from section \ref{strategy}. To test whether
 eq. (\ref{tpphi}) or eq. (\ref{threepoint2})  has some relevance 
 for our data   we will consider whether the following quantities:
\beq  {F^0(\vec p,\vec q ;t)}{\left [
 \,p^0 f_\pi \,\frac{ e^{- ( E_\pi/2 + E_S)\,t}}{2 E_S}\,
\Phi_\pi(1/2)\right ]^{-1}}
\label{ratiopart}
\eeq
and
\beq F^0(\vec p,\vec q ;t)\left[e^{- E_{\widetilde\pi} \,t}\right]^{-1}
 \label{ratiosp}
\eeq
are constant in time for some time interval.

\begin{figure}[hbt]
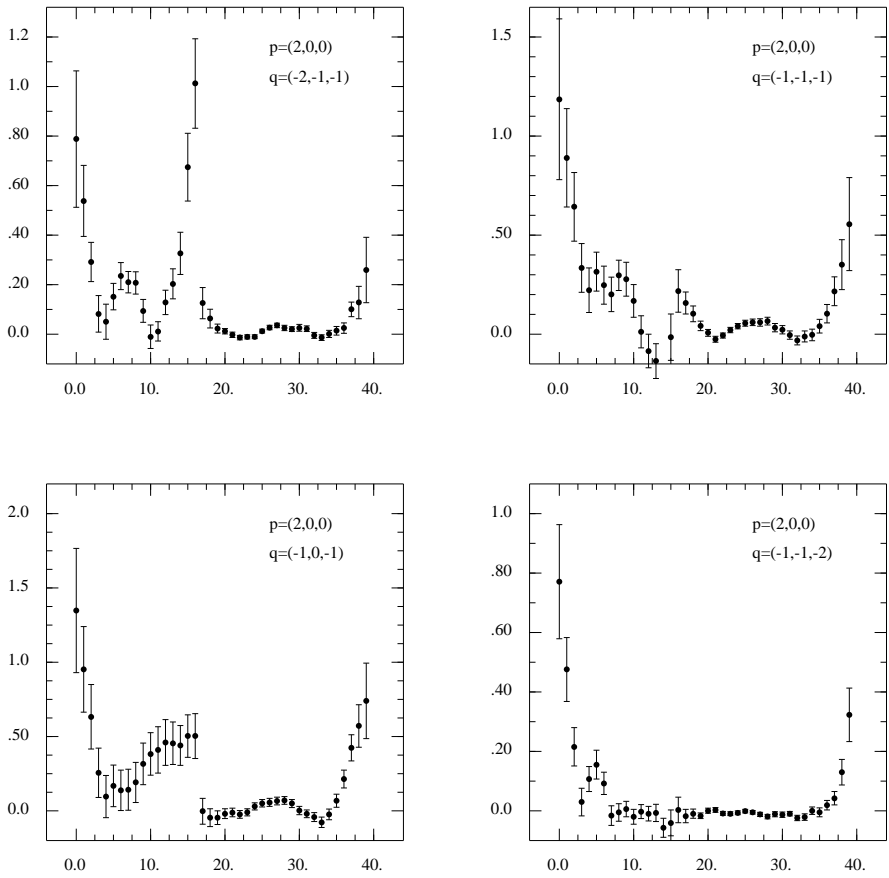

\hspace*{-1.3cm}
\begin{center}
\begin{tabular}{ll}
\epsfxsize6.0cm\epsffile{trois-p5_full_200_-2-1-1.eps} & 
\epsfxsize6.0cm\epsffile{trois-p5_full_200_-1-1-1.eps} \\
\epsfxsize6.0cm\epsffile{trois-p5_full_200_-10-1.eps} &
\epsfxsize6.0cm\epsffile{trois-p5_full_200_-1-1-2.eps} 
\end{tabular}
\end{center}
\caption{\small We plot $F^0(\vec{p},\vec{q};t)$,  normalized by a constant 
(divided by the pion propagator with $\vec p=(4\pi/L,0,0)$ from the fixed time 
$t_\pi$ to $0$)  versus 
the running time for momenta indicated on the plots
 using the lightest quarks ($\kappa=0.1339 $) and the 
 lightest ``squark'' ($\kappa_S=0.1431$)..}  
  \label{exemples}
\end{figure}

Before that, it is instructive to have a look at the numerators 
$F^0(\vec p,\vec q ;t)$.  In  figure \ref{exemples} we have plotted as an
illustrative  example the three-point function  for $\vec p_\pi= {2 \pi}/L\ (2,0,0)$
 and various vectors $\vec q$. We observe a very striking feature akin to an
oscillating behavior.  We do not claim to understand fully this shape. However
since, 
in section \ref{principle}, a rationale was  elaborated to  describe the expected
partonic  behavior which may show up at small time, we will  focus  
from now on on this time interval. \\
The very rapid drop observed at small time, i.e. $t\in [0,3-4]$ is present for
all  values of $\vec q$. {\it We will test the hypothesis that this rapid drop is
due to a partonic signal} assuming that the hadronic behaviour sets in for larger
times.  The typical shape in fig. \ref{exemples} might  suggest a negative
interference between the small time regime and the later one, leading to a
vanishing amplitude  around $t = 4$. 
We do not understand the origin of the
latter, which is beyond the scope of this work focused on the small-time drop.  
It is
noticeable that the statistical errors for this time range are small enough to 
exhibit a signal while the two-point function for the corresponding pion 
propagation time and the same pion momentum  is extremely noisy.

\paragraph{
 Searching for plateaus at small times: } A plateau
of eq. (\ref{ratiopart}) would indicate a partonic-like behavior, while a plateau
of eq. (\ref{ratiosp}) would sign a pionino. 
 We will
compute $E_\pi$ and $E_{\widetilde\pi}$ from the measured  pion and pionino rest
masses, see tab. \ref{tab:deux_points}, and the formula $E=\sqrt{m^2+p^2}$. We
prefer this to the direct use of the  measured energies for non zero momentum,
reported in tab. \ref{tab:deux_points}, because the latter are noisier than the
rest masses for $\vec p={2\pi\over L}(2,0,0)$.

The energy $E_S$ has been taken via eq. (\ref{ES})  assuming  two possible
masses $m_S$ for $\kappa_S=0.1431$. As already mentioned, for  $\kappa_S>0.1431$
the calculation of $D(x,0)$ from eq. (\ref{propagateur}) fails indicating the
presence of small eigenvalues, i.e. that $m_S$ is small.    
Besides  considering
a massless scalar parton ($m_S=0$), we have also considered the value
 $m_S=0.22$  in lattice units, which corresponds to the
scalar-scalar bound state mass (divided by two). It would be tempting  to fit
$m_S$ from the results yielding  the flattest plateau, but it turned out to be too
difficult to disentangle the effect of $m_S$ on the plateau from other effects
which will be discussed later.

\begin{figure}[hbt]
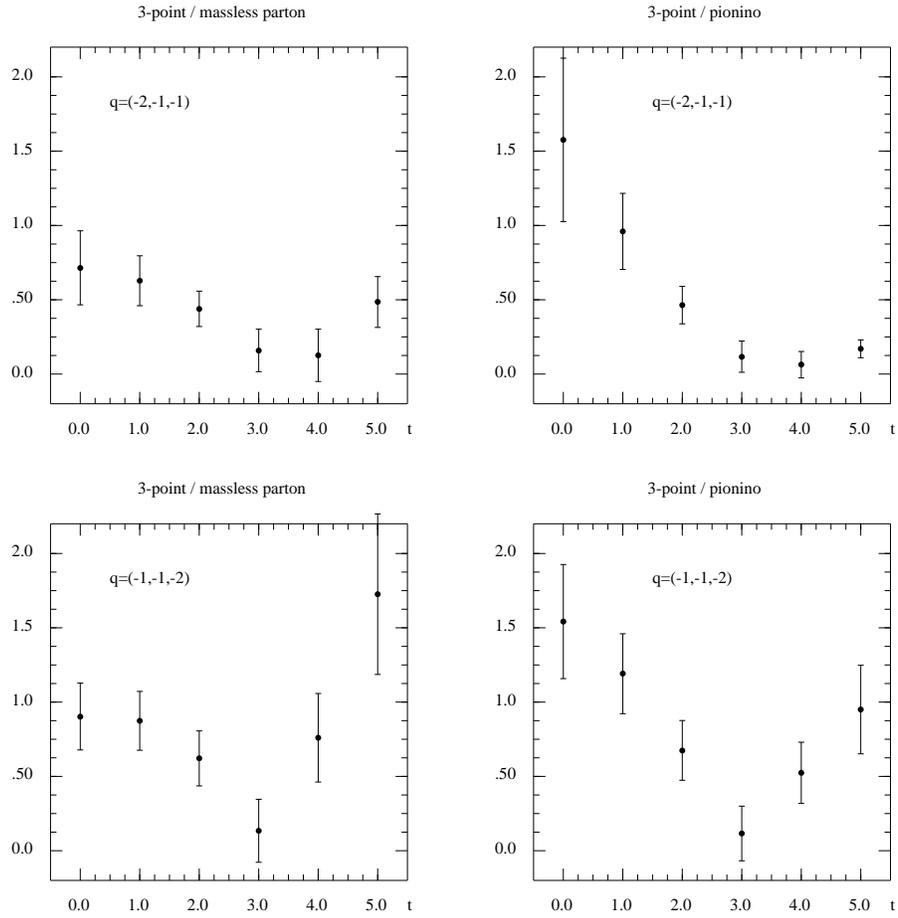

\hspace*{-1.3cm}
\begin{center}
\begin{tabular}{ll}
\epsfxsize6.0cm\epsffile{3-point-parton0_-2-1-1.eps} & 
\epsfxsize6.0cm\epsffile{3-point-pionino_-2-1-1.eps} \\
\epsfxsize6.0cm\epsffile{3-point-parton0_-1-1-2.eps} & 
\epsfxsize6.0cm\epsffile{3-point-pionino_-1-1-2.eps} \\
\end{tabular}
\end{center}
\caption{\small Ratios of eqs. (\ref{ratiopart}) (left) and
(\ref{ratiosp}) (right)  for momenta indicated on the plots
 using the lightest quarks ($\kappa=0.1339 $) and the 
 lightest ``squark'' ($\kappa_S=0.1431$). }  
  \label{fig:rapports}
  \end{figure}

In fig. \ref{fig:rapports} we show two examples of ratios corresponding to eqs.
(\ref{ratiopart}) (left) and (\ref{ratiosp}) (right) at small time. 
In the light of the
discussion  in subsection \ref{derivation}, we have chosen as illustrative the
following kinematics:  $L\vec q/(2\pi) = (-2,-1,-1)$ and $L\vec q/(2\pi) =
(-1,-1,-2)$, both for  $L\vec p/(2\pi) = (2,0,0)$. It is clearly seen that 
that the plots to the right (\ref{ratiosp}) are utterly incompatible with a
plateau, thus discarding a pionino interpretation at small time. The plots to the
left might show some indication of plateaus but they deserve some discussion. 
The
signal decreases from a maximum at time $0$ to reach a value  compatible with $0$
at a time $3-4$. This happens not only for these two examples but is a general 
pattern for all the kinematics considered. This cancellation has already been seen
on the numerators of eq. (\ref{ratiopart}) in figs. \ref{exemples}. 
We have argued that it is motivated by destructive interferences 
which  generate an overdecreasing of the numerators in eq.  (\ref{ratiopart})
 with respect to the denominators.
 The signals vanish as soon as $t=3-4$, restricting the range where  plateaus
might be seen to a very short time interval~\footnote{One may worry about contact terms or other lattice artifatcs that might 
spoil the analysis around $t=0$, this will be discussed in the conclusion.}
 around $t=0$.

Anyhow, the most restrictive of the constraints relative to $t$, summarized at
 the end of subsection
\ref{derivation}, i.e.  $t \ll  \, E_S/(\Lqcd |\vec{q}_\perp|)$,  
amounts, for a
massless scalar parton, with our lattice set-up  and the value $u=1/2$,  
to the condition 
\beq
t \ll {a^{-1}\over\Lambda_{\rm QCD}}{\sqrt{(p_x/2+q_x)^2+q_\perp^2}
\over |q_\perp|} \sim 5 \, \label{cdt1}
\eeq 
where for $\Lambda_{\rm QCD}$ we have taken a typical quark transverse momentum 
of 400 MeV within a hadron.  
This constraint does not allow to use larger time domains than the one just discussed.

We will now  go on confronting the slopes on this  
small time interval to the 
theoretical prediction of a plateau for eq. (\ref{ratiopart}), postponing the
maybe more convincing comparative study of the values of $F^0(\vec p,\vec q;0)$.

We perform a systematic study over a larger set of three 
point Green functions defined such that: 
$L \vec p/2\pi = (2,0,0)$,  $(L \vec q/2\pi)^2 \le 4$
and $(L \ (\vec q+\vec p)/2\pi)^2 \le 6$. These limitations on
the norm of the momenta are meant to avoid too noisy results. 
 On the other hand, the constraint $E_S \gg \Lambda_{\rm QCD}$ (see subsection
 \ref{derivation}) translates into the lower bound:
 
\beq
\frac{L}{\pi}\sqrt{\left (\frac{p_x}{2}+ q_x\right )^2+q_\perp^2 } \gg 1
\label{cdt2}
\eeq

\begin{figure}[hbt]
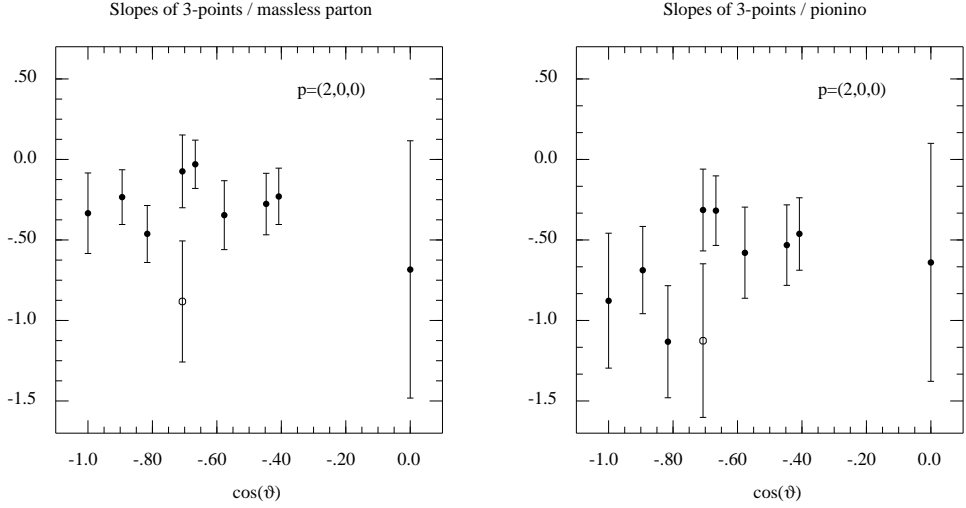

\hspace*{-1.3cm}
\begin{center}
\begin{tabular}{ll}
\epsfxsize6.5cm\epsffile{pente_part0.eps} & 
\epsfxsize6.5cm\epsffile{pente_pionino.eps} \\
\end{tabular}
\end{center}
\caption{\small Slope of the ratios on the time interval
t=0,4 for formulae (\ref{ratiopart}) (left) and
(\ref{ratiosp}) (right) for different values of $\vec q$ and for $\vec p =
{2\pi\over L}(200)$ with a massless scalar parton. The horizontal axis
is the cosine of the angle between vectors $\vec p$ and $\vec q$.  
}  
  \label{fig:ratios}
  \end{figure}

For this set of data we measure the slope of the ratios in eqs. 
(\ref{ratiopart}) and (\ref{ratiosp}) for the time intervals $t=0,3$ and
$t=0,4$.  For the latter range, the results are presented  in
fig. \ref{fig:ratios}: the ratios of eqs. (\ref{ratiopart}) and (\ref{ratiosp})
are presented for commodity as a function  of the cosine of the angle between $\vec p$ and
$\vec q$, which we will from now on refer to as $\cos \theta_{pq}$. 

We have eliminated from the analysis the data with 
$L\vec q/(2\pi) = (-1,0,0)$ for
which the scalar parton is at rest ($p_x/2+ q_x=0$) and  thus violates the
condition \eq{cdt2}. The  data with white circles on the plots 
correspond to  $L\vec q/(2\pi) = (-1,0,-1)$ which is marginal for both conditions 
eqs. (\ref{cdt1},\ref{cdt2}). 
It could be noticed that the back-to-back points  $L\vec q/(2\pi) = (-2,0,0)$ do
not raise problems as a result  of the discretization of partonic momenta. Indeed,
since $u=1/2$, $u \vec p +\vec q$ never vanishes contrarily to the continuum case
 discussed in  section \ref{derivation}. More generally, the majority of the points with 
$\cos \theta_{pq}$ close to -1 are not excluded for the same reason. 

Comparing both plots in fig. \ref{fig:ratios}, it is evident that the  partonic
slopes (left) are much closer to zero than the hadronic ones (right).
Nevertheless, the partonic slopes show a general
tendency to be negative (see tab. \ref{tab:trois_points}) 
which can be traced back to the vanishing around $t=3-4$.  
The white circle points show a lesser improvement of the partonic data as compared to
the hadronic ones as conjectured just above.

\begin{table}[hbt]
\begin{center}
\begin{tabular}{|c|c|c|c|c|}
\hline
 model &time slice & $\chi^2$/d.o.f & average slope \\ \hline
 pionino & 0-4 & 4.1 & -0.56(18) \\
 partons $m_S=0$& 0-4 &1.9  &-0.26(13) \\
 partons $m_S=0.22$ & 0-4&0.92  &-0.23(13) \\
 pionino & 0-3 & 8.3 & -0.82(9) \\
 partons $m_S=0$ & 0-3&3.7  &-0.39(13) \\
 partons $m_S=0.22$ & 0-3& 2.2 &-0.36(13) \\

\hline
\end{tabular} \par
\vskip 5 truemm
\end{center}
\caption{\small Average slopes (and $\chi^2/$d.o.f for a vanishing slope) of the
expression appearing in eqs. (\ref{ratiopart}) 
and (\ref{ratiosp}) for two time slices and two parton masses. 
It is seen that the parton mass does not play 
a very important role. The difference between the two time 
slices is due to the zero of $F^0$ discussed in the text.}
\label{tab:trois_points}
\end{table}
\vskip 5 truemm

 The slopes given in table \ref{tab:trois_points} are the averages over our
  set of momenta $\vec q$ (excluding the momentum corresponding to the white circle).
 We have kept the mass of the scalar parton between 0 and half the mass of the
 scalar-scalar bound state (see tab. \ref{tab:deux_points}). 
 The resulting slopes do not depend significantly on the latter mass.
It can also be seen that the slopes are quite similar for time-slices $[0,3]$ and 
 $[0,4]$.

\paragraph{
 Comparing three-point functions at $t=0$: } 

Equation  (\ref{tpphi}) predicts two main features of the partonic behaviour:\\
i) the exponential time evolution \\
ii) the following amplitude at $t=0$ 
\beq 
F^0(\vec p,\vec q;t=0) \propto \frac{\Phi_\pi(u=1/2)}{2E_S(u=1/2)}.
\label{ampt0}
\eeq

\noindent The begining of this section was devoted to 
the time evolution. Let us now focus on the amplitude \eq{ampt0}.

\begin{figure}[hbt]
\hspace*{-1.3cm}
\begin{center}
\begin{tabular}{ll}
\epsfxsize6.5cm\epsffile{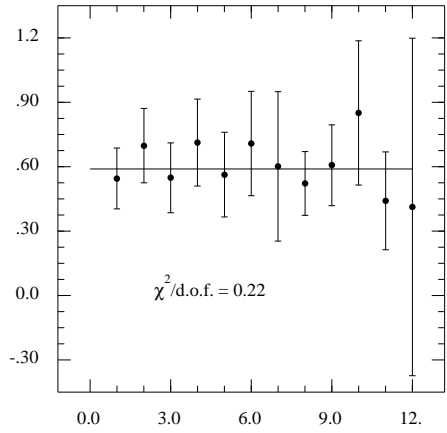} 
\end{tabular}
\end{center}
\caption{\small Values of $E_S(1/2) F^0(\vec p,\vec q;0)$ 
normalized as fig. \ref{exemples} for $\vec p=(4\pi/L,0,0)$ 
and our full set of $\vec q$ (labeled from 1 to 12 on the horizontal axis). 
 The data show the expected constancy 
around the average represented by the horizontal line.}  
  \label{fig:val0}
  \end{figure}

The  plot in fig. \ref{fig:val0} shows for our set of momenta $\vec q$ 
the product $E_S(1/2) F^0(\vec p,\vec q;0)$ which is expected to
be constant from \eq{ampt0}. 
$E_S$ is computed  from eq. (\ref{ES}) with  a massless scalar parton.
 The plotted ratio is indeed strikingly constant: the $\chi^2/{\it d.o.f.}$ for the fit
to a constant ratio is $0.22$. This expected constancy of a large set of numbers, 
which are significantly different from zero,  yields an amazing support to a 
partonic interpretation of these data. We cannot figure out any  other explanation 
for this feature. Indeed, one might fear that the observed constancy of 
$E_S(1/2) F^0(\vec p,\vec q;0)$ is simply due to some contact term producing 
a $\vec q$ 
independent $F^0(\vec p,\vec q;0)$ combined with a small dependence of 
$E_S(1/2)$ on $\vec q$. To
consider this we have tried a fit with $F^0(\vec p,\vec q;0)=$~constant,  
which gives $\chi^2/{\it d.o.f.}= 0.72$ ,larger than the previously found  0.22,
 although still 
smaller than 1.
We would thus rather believe, in agreement with  the partonic interpretation,
that the small variation of 
$F^0(\vec p,\vec q;0)$ is a consequence of the constancy of 
$E_S(1/2) F^0(\vec p,\vec q;0)$ and a small variation of $E_S(1/2)$.
As a check, we have tested the constancy of $F^0(\vec p,\vec q;0)$   for 
$p=2\pi/L (1,0,0)$, which is  not expect to follow \eq{ampt0} while contact
terms have no reason to be absent~\footnote{We did not 
check the constancy of $E_S(1/2) F^0(\vec p,\vec q;0)$ in this case since  $u=1/2$
is forbidden in the case $p=2\pi/L (1,0,0)$.}. We find $\chi^2/{\it d.o.f.}= 2.7$
which further supports the partonic interpretation of the constancy 
$E_S(1/2) F^0(\vec p,\vec q;0)$ for $p=2\pi/L (2,0,0)$.

\section{Discussion and Conclusion}
\label{conclusion}

\parin We have performed the first tentative application 
of a new proposal \cite{aglieti} to compute the pion LCWF.
This proposal was to compute the pion to vacuum matrix element
of a non-local operator, namely 
the propagator of a scalar particle which has the color quantum numbers
of a quark. For convenience, we call it a ``squark". 
This resulting matrix element is gauge invariant. To 
exhibit the partonic structure of the pion a large momentum $\vec q$ is added
to the scalar propagator.   

We have shown that, provided the pion has a large enough
momentum $\vec p$, provided  that the squark has a large enough energy,
and provided the propagation time of the scalar object is short enough
(end of subsection \ref{derivation}),
the  above mentioned matrix element is dominated by a contribution
from the pion LCWF. A measure of this matrix element can then 
provide informations on the LCWF.

 A necessary first step is the computation of the two point Green functions
of quark (squark) - quark (squark) bound states. The new states,
which contain at least one squark, show a behaviour quite similar to standard
hadrons, they show nice exponential time dependence, fig. \ref{fig:twopoint},
they verify Einstein spectral law, fig. \ref{fig:einstein}, and the masses 
decrease with increasing $\kappa_S$ i.e. decreasing squark bare mass.

We have then analyzed the three point Green functions for a large set of  
pion momenta $\vec p$ and transfers $\vec q$. The scalar parton has a momentum 
$u \vec p +\vec q$ where $u\in[0,1]$ is the fraction of pion momentum carried by the active quark.
The discretization due to the finite volume implies a discretization of the fraction $u$.
In our set, only the momentum $\vec p={2\pi\over L}(2,0,0)$ allows for 
$u\ne 0,1$ (where the LCWF vanishes), namely $u=1/2$. 

 We focused the analysis on small times
($t\in [0,3-4]$) according to the formulae (\ref{ratiopart}) and (\ref{ratiosp})
which express respectively the hypothesis of a partonic behaviour
of the squark and the spectator quark during this small time interval,
or, on the contrary, the hypothesis of a precocious confinement
of the squark and the spectator quark into a 
hadronic-like bound state. The correct hypothesis should show up as a 
plateau in time. 

Our data clearly favor the partonic behaviour at small time: the observed rapid drop 
of the Green function is expected from a partonic picture, while a hadronic picture predicts 
slower decrease.   
The analysis is however made delicate due to an observed vanishing of the Green function 
around $t=3-4$ which might be due to a destructive interference. The resulting analysis domain 
is very short and close to zero. 
 This might induce the objection that we cannot disentangle our signals from lattice 
 artifacts such as contact terms, etc. 

Nevertheless, a second series of tests has confirmed our feeling that a real partonic 
signal shows up: all the Green functions at $t=0$ for our set of values 
of $\vec q$ verify the prediction, eq. (\ref{ampt0}), of the partonic model (up to one unknown constant) 
in an amazing manner. It is difficult to figure out how a lattice artifact
  could mimic this behaviour for so many data.

This work aimed mainly at testing the viability of this program.
We believe that the answer is positive. 
The fact that we could argue rather firmly that we see 
a partonic signal is encouraging, obtained on a small lattice, with
a rather large lattice spacing, and ``large'' momenta which are indeed
not so large !

 In order to progress we first need to settle the question of 
 possible lattice artifacts. To that aim, it would be necessary to change the 
 lattice parameters and mainly $a$ and to run a larger set of momenta.
 This would furthermore allow to reach other values of $u$ than 1/2 and provide 
 an idea about the shape of the LCWF.
 This program implies the use of a larger volume which 
 would also hopefully reduce the noise of large
momenta Green function.

 A recent work by S. Dalley based on a Hamiltonian formulation of
  QCD on a lattice \cite{Dalley} 
 presents an interesting analysis of the LCWF.  
 This new method is very
 promising although it presents some difficulties as stated 
 by the author. It is of course too
 early to perform a detailed comparison of the Lagrangian formulation used here
  and the Hamiltonian one. Both need 
 to be followed.
\section{Acknowledgments} 

\parin We are specially grateful to Guido Martinelli and Damir Becirevic 
for the discussions that initiated
this work. We thank 
 Gregori Korchemsky and Claude Roiesnel for very instructive discussions. 
  J. R-Q is indebted
to Spanish  Fundaci\'on Ram\'on Areces for financial support. These calculations
were performed on the QUADRICS QH1 located  in the Centre de Ressources
Informatiques (Paris-sud, Orsay) and purchased thanks to a funding from the
Minist\`ere de l'Education Nationale and the CNRS.
This work is supported in part by European Unions Human Potential Program under contract
HPRN-CT-2000-00145 Hadrons Lattice QCD.


\begin{thebibliography}{9}
{\small
\bibitem{brodsky}
S. J.~Brodsky, Y.~Frishman, G. P.~Lepage and C.~Sachrajda,
{\em Phys. Lett.} {\bf 91B}, 239 (1980);
S. J.~Brodsky, Y.~Frishman and G. P.~Lepage,
{\em Phys. Lett.} {\bf 167B}, 347 (1986);
  S. J. Brodsky and G. Peter Lepage, "Exclusive processes in Quantum
chromodynamics", contribution to
"Perturbative Quantum Chromodynamics"; Ed. by A.H. Mueller, World
Scientific Publising Co. (1990);
SLAC-PUB-4947.
\bibitem{stan1} S.J. Brodsky, hep-ph/9908456, SLAC-PUB-8235.
\bibitem{farrar}  S.J. Brodsky and G.R. Farrar,  Phys. Rev. Lett. {\bf 31}, 
1153 (1973).
\bibitem{efremov} A.V. Efremov and A.V. Radyushkin, Theor. Math. Phys. {\bf 42}, 97
(1980).
\bibitem{bertsch} G. Bertsch, S.J. Brodsky, A.S. Goldhaber, and J. Gunion,
               Phys. Rev. Lett. {\bf 47}, 297 (1981). 
\bibitem{CZ} V.L. Chernyak and A.R. Zhitnitsky, Phys. Rep. 112 (1984) 173.
\bibitem{BF} V.~M.~Braun and I.~E.~Filyanov, Z. Phys. C {\bf 44} (1998) 157.
\bibitem{stesto} G. Sterman and P. Stoler, Ann. Rev. Nuc. Part. Sci.      
{\bf 43}, 193 (1997), hep-ph/9708370.
\bibitem{cleo}CLEO Collaboration, J. Gronberg {\it et al.} Phys. Rev. 
           D{\bf 57}, 33 (1998).
\bibitem{rad} P. Kroll and M. Raulfs, Phys. Lett. B{\bf 387}, 848 (1996);
I.V. Musatov and A.V. Radyushkin, Phys. Rev. D{\bf 56}, 2713 (1997); 
A.~Schmedding, O.~Yakovlev, Phys. Rev. D{\bf 62}(2000) 116002;
V.~M.~Braun, A.~Khodjamirian and M.~Maul, Phys. Rev. D{\bf 61}(2000) 073004.
\bibitem{E791}E.M. Aitala et al. (Fermilab E791 coll.), hep-ex:0010043.
\bibitem{THE791} V.M. Braun et al. hep-ph/0103275; V. Chernyak hep-ph/0103295.
\bibitem{sac1} G.~Martinelli and C.T.~Sachrajda,  Phys.~Lett. 190B (1987) 151
and  Nucl.~Phys. B316 (1989) 305.  
\bibitem{aglieti} U. Aglieti, M. Ciuchini, G. Corb\`o, E. Franco, 
G. Martinelli, L. Silvestrini, Phys. Lett. B441 (1998) 371.
\bibitem{Dalley} S. Dalley, hep-ph/0101318.}
\end{thebibliography}
\end{document}